 \documentclass[final,5p,times,twocolumn,authoryear]{elsarticle}

\usepackage{amssymb}
\usepackage{lipsum}
\usepackage{amsthm}
\usepackage[colorlinks=true,allcolors=blue]{hyperref}

\usepackage{enumitem}
\setitemize{noitemsep}

\usepackage[labelfont=bf]{caption}

\usepackage[linesnumbered,ruled,vlined]{algorithm2e}
\DontPrintSemicolon

\SetKwComment{Comment}{\color{green!50!black}// }{}

\SetKwProg{Function}{function}{}{}

\usepackage{multirow}
\usepackage[table,xcdraw]{xcolor}

\journal{Astronomy $\&$ Computing}

\begin{document}

\begin{frontmatter}



\title{Towards the exploration of astrophysical datacubes through volumetric rendering within CARTA}


\author[iaa]{I. Labadie-García}
\ead{ixakalab@iaa.es}
\author[idia]{A. Pińska}
\author[idia]{A. Comrie}
\author[idia]{M. van Zyl}
\author[iaa]{M. Parra-Royón}
\author[iaa]{S. Sánchez-Expósito}
\author[iaa]{L. Verdes-Montenegro}
\author[iaa]{J. Garrido}

\affiliation[iaa]{organization={Instituto de Astrofísica de Andalucía},%Department and Organization
            addressline={Glorieta de la Astronomía s/n}, 
            city={Granada},
            postcode={18008},
            country={Spain}}
            
\affiliation[idia]{organization={Inter-University Institute for Data Intensive Astronomy},
            adressline={University of Cape Town, Rondebosch},
            city={Cape Town},
            postcode={7701},
            country={South Africa}}

\begin{abstract}

Data products from astrophysical observations may contain three or more dimensions: most commonly two spatial dimensions and a third spectral dimension, and possibly additional axes such as polarisation or time. Visualising three-dimensional image data is particularly useful for analysing the kinematics and evolution of various objects; however, the large scale of data produced by next-generation observatories and the complexity of displaying 3D data on a 2D screen make visualisation challenging. In this article, we present a scalable and interactive 3D visualisation approach within the astronomy software CARTA. We implemented a prototype 3D rendering widget with ThreeJS and known techniques to improve efficiency (e.g. compression, downsampling) and found that large datacubes up to 128 GB can be visualised locally with a loading time of a few minutes. We analysed the performance in the frontend and in the backend, finding that the bottleneck of our implementation lies in the frontend. This is due to storing the data in a single array in the frontend, unlike in the backend, where it is accessed and sent in blocks. This study paves the way for making 3D visualisation of large datacubes available to the community by implementing a 3D rendering widget in a future release of CARTA.

\end{abstract}




\end{frontmatter}
\tableofcontents



\section{Introduction}\label{introduction}

Data visualisation has consistently been used throughout history for scientific studies. Analysing data by eye helps humans understand the physical phenomena at play and recognise patterns that would otherwise remain hidden using numerical analysis; astrophysics is no exception. While numerical methods produce concrete and definite results, in ambiguous and uncertain cases, visual or auditive interpretation can be essential. In practice, even though automatic processes are used to relieve the workload of analysing extensive datasets, critical discoveries require careful visual inspection. Astronomical data spans a range of dimensions---1D plots, such as spectra or histograms; 2D images, such as photometric observations; and three-dimensional (3D) datacubes, such as spectral cubes or cosmological simulations.

Among these, datacubes are very valuable for studying the morphology and structure of astrophysical sources. By representing data in 3D, astronomers can more effectively interpret spatial and kinematic properties of objects, such as galaxies or molecular clouds. This type of data is mostly produced by radio telescopes and integral field units (IFU), which capture both spatial and spectral information. This results in inherently large data volumes that can be challenging to visualise. The increasing sensitivity and resolution of modern observatories, e.g., ALMA \citep{ALMA2009}, MeerKAT \citep{meerkat2016}, ASKAP \citep{askap2021}, JWST \citep{jwst2006}, or MUSE \citep{muse2010}, have increased data sizes to GB or even TB scales. Data size will continue growing with next-generation observatories like SKAO\footnote{\url{https://www.skao.int/en}} or ngVLA\footnote{\url{https://ngvla.nrao.edu/}}, which are expected to generate unprecedented data rates and volumes.

The astrophysics community is making a large effort to prepare for this task. For example, SKAO is developing a network of computing facilities distributed around the world (SRCNet; \citealp{srcnet0.1_devplan}) to store and give access to the data. Since data will be too large to download on standard desktop computers, the SRCNet will also need to provide processing, analysis and visualisation services.

The topic of visualisation in astronomy, particularly in the context of large-scale and multidimensional data, has been reviewed widely in recent years (e.g., \citealp{hassanfluke2011_3Dreview}, \citealp{Lan2021_3Dreview}). Some noteworthy tools include: VisIVO \citep{visivo2015}, a framework designed for the Virtual Observatory (VO) to visualise multidimensional data through various methods; ViSL3D \citep{visl3d2025}, a Python library made to create 3D iso-surface visualisations through a VO science archive; iDaVIE-v \citep{idavie2024}, an innovative virtual reality software for 3D astronomical data, although it has broader applications; and CARTA \citep{carta}---the focus of this work---which will be described in detail in Sec. \ref{Architecture}. These tools approach the issue of big data by implementing a client-server architecture; that is, they take advantage of high-performance remote servers to compute the bulk of operations, leaving only rendering and interaction to the client, reducing hardware requirements for users.

%To address these challenges, we previously developed ViSL3D, a VO-compatible tool focused on scalable 3D visualisation of iso-surfaces, but other approaches can be equally valid.
Despite these advancements, 3D visualisation in astronomy remains limited. Many existing tools either lack native 3D capabilities, are not sufficiently interactive, or cannot handle large datasets efficiently because of memory and processing constraints. In this article, we introduce a 3D rendering widget within CARTA, designed to address these challenges. We have used ThreeJS\footnote{\url{https://threejs.org/}} to implement a ray marching algorithm that enables interactive exploration of spectral datacubes and similar 3D data. We use techniques to optimise data-handling from the request of the visualisation to the rendering. 

\subsection{Architecture}\label{Architecture}

The Cube Analysis and Rendering Tool for Astronomy, CARTA, is a data visualisation and analysis application that addresses the challenges of big data in astronomy. \cite{hassanfluke2011_3Dreview} argue that remote and distributed platforms are most efficient to handle large datasets. First, the data will be too large to download to a personal computer or too large to manipulate. Accessing the data on a remote server would lessen this concern. Additionally, processing the data in a distributed way, using many computing nodes, can improve the performance of data-intensive tasks.

CARTA adopts a client–server architecture to achieve a balance between computational efficiency and user interactivity. This way, data is stored and processed on a remote server, where analysis calculations---such as generating moment maps, calculating histograms, or producing cutouts---can be done efficiently. Only the results of these calculations are transmitted to the user interface for visualisation, significantly reducing data transfer requirements. For flexibility, CARTA also supports a local deployment mode, where both the client and server run on the same machine. This configuration retains the same functional features as the remote version but with limited computational scalability.

Another major factor contributing to the performance of CARTA is the use of a hierarchical data format optimised for rapid access: HDF5-IDIA \citep{hdf52020}. This format includes precomputed datasets such as copies of the data with permuted axes, downsampled \textit{mipmaps}, histograms, and statistics. This structure enables faster access to subsets of data and reduces redundant computation during visualisation or analysis tasks. \cite{hdf52020} offer more details on the schema and show that using this format accelerates the tasks to be performed.  CARTA supports this format in addition to FITS and other common astronomical data formats. A conversion tool for FITS is available online\footnote{\url{https://github.com/CARTAvis/fits2idia}}.

\begin{figure*}
    \centering
    \includegraphics[width=0.7\linewidth]{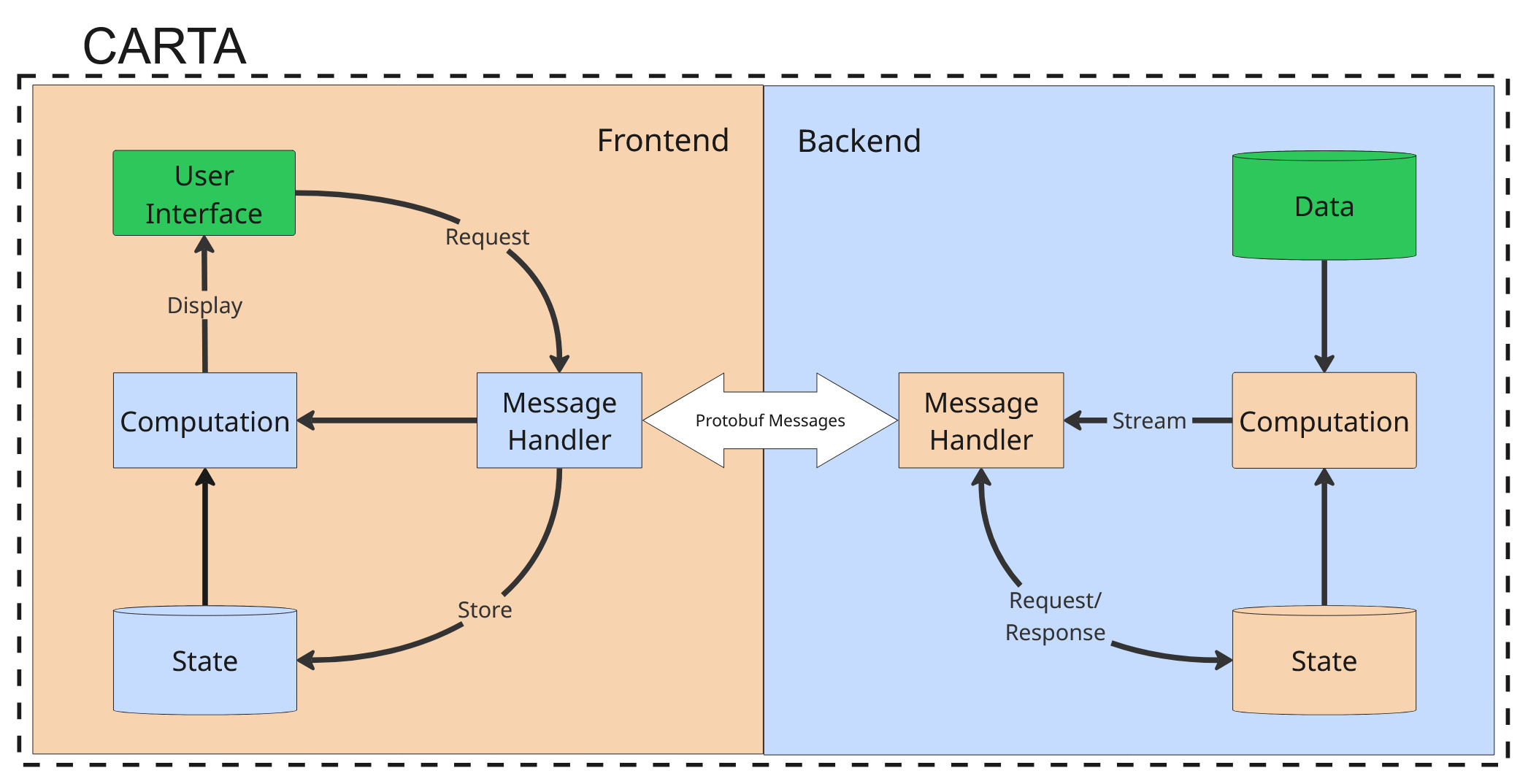}
    \caption{Simplified diagram of the architecture of CARTA. It represents components of the software and how they are connected.}
    \label{fig:architecture}
\end{figure*}

Figure \ref{fig:architecture} is a simplified diagram showing the main components of CARTA, demonstrating the workflow of standard widgets or functions such as spectral profiling, creating histograms or generating position-velocity images. The user interface is responsible for displaying the data and components as well as allowing the user to request information and introduce parameters. The message handler creates the request and communicates with the backend through protobuf messages. The result returned by the backend may require further frontend-side computation before it can be displayed to the user: for example, numeric raster image values are mapped to colours. The frontend caches part of the data in its object state to avoid repeating the same requests unnecessarily. When a request arrives in the backend, it is managed by a handler component, which dispatches it to the appropriate backend computation component. These components update backend object states and read data from disk as needed, and optionally compute other data products, for example spectra or moment maps. Data are sent back to the frontend through the backend message handler: small data products are typically returned in a response to the request, but larger data may be streamed as an independent sequence of messages, avoiding bottlenecks and improving efficiency.

The user interface of CARTA is web-based, aligning with the approach taken by current tools (e.g., Aladin, \citealp{aladin2022} or VisIVO) because 1) users can access it from any modern web browser, independently of the operating system; 2) it can be seamlessly integrated with remote and cloud resources; and 3) it can incorporate established visualisation or processing libraries: for example, to visualise 2D images and plots, CARTA imports Charts.js\footnote{\url{https://www.chartjs.org/}} and Plotly\footnote{\url{https://plotly.com/}}. The backend of CARTA is primarily implemented in C++, optimised for high-performance computation and memory management, while the frontend is developed in TypeScript.

\section{3D rendering}\label{3DRendering}

To enable 3D visualisations within CARTA, we have included ThreeJS\footnote{\url{https://threejs.org/}}, a JavaScript (JS) library designed to render and manipulate 3D content in web environments. ThreeJS directly handles scenes, materials, lighting, textures and other elements with WebGL\footnote{\url{https://www.khronos.org/webgl/}}. In general, 3D rendering tools employ one of two techniques: volume rendering or texture-based methods.

Volume rendering treats the entire datacube as a continuous 3D field, representing every voxel without extracting geometric surfaces. The visualisation is created with ray marching or ray casting, which involves directing rays through the volume and extracting values along the way; these values are mapped to the colour and opacity to be displayed. The ray marching can be done with regular step sizes, adapting the step depending on parameters such as the gradient, or with other conditions to improve performance.
On the other hand, texture-based methods extract geometric information from the data and map it into polygonal meshes or textures. In most cases, the information extracted from datacubes takes the form of iso-surfaces, that is, surfaces formed by voxels of equal value. Iso-surfaces can be displayed with different colours and opacities to enable distinguishing them and seeing through. Libraries such as VTK.js\footnote{\url{https://kitware.github.io/vtk-js/index.html}} or X3DOM\footnote{\url{https://www.x3dom.org/}} were considered for potential use, especially with iso-surface representations in mind, but ThreeJS was selected for its simplicity and flexibility.

Visualising iso-surfaces requires significant pre-processing to compute the surfaces prior to rendering, which can be computationally expensive. One option is to calculate iso-surfaces in the frontend using libraries that include functions for that purpose. The other option is to perform the computation in the backend, implementing a code such as the Marching Cubes Algorithm \citep{marching2003}, to calculate the surfaces. Then, only the surfaces would be sent to the frontend for rendering. The latter is the most favoured approach for optimisation. Using the backend for computation is a similar strategy to the one used by the contour widget, but with an additional dimension. This approach would also require a histogram to choose the levels to represent. Volume rendering with ray marching offers a more straightforward implementation, as it does not require this additional effort. Furthermore, it provides more flexibility since iso-surfaces can be effectively reproduced using certain colour transfer functions. Ray marching has proven to faithfully represent astrophysical objects including diffuse emission; e.g., iDaVIE-v, DS9 \citep{ds92003}, or VisIVO; therefore it has been chosen as the first method to be integrated into CARTA.

\begin{figure}[ht]
    \centering
    \includegraphics[width=0.45\textwidth]{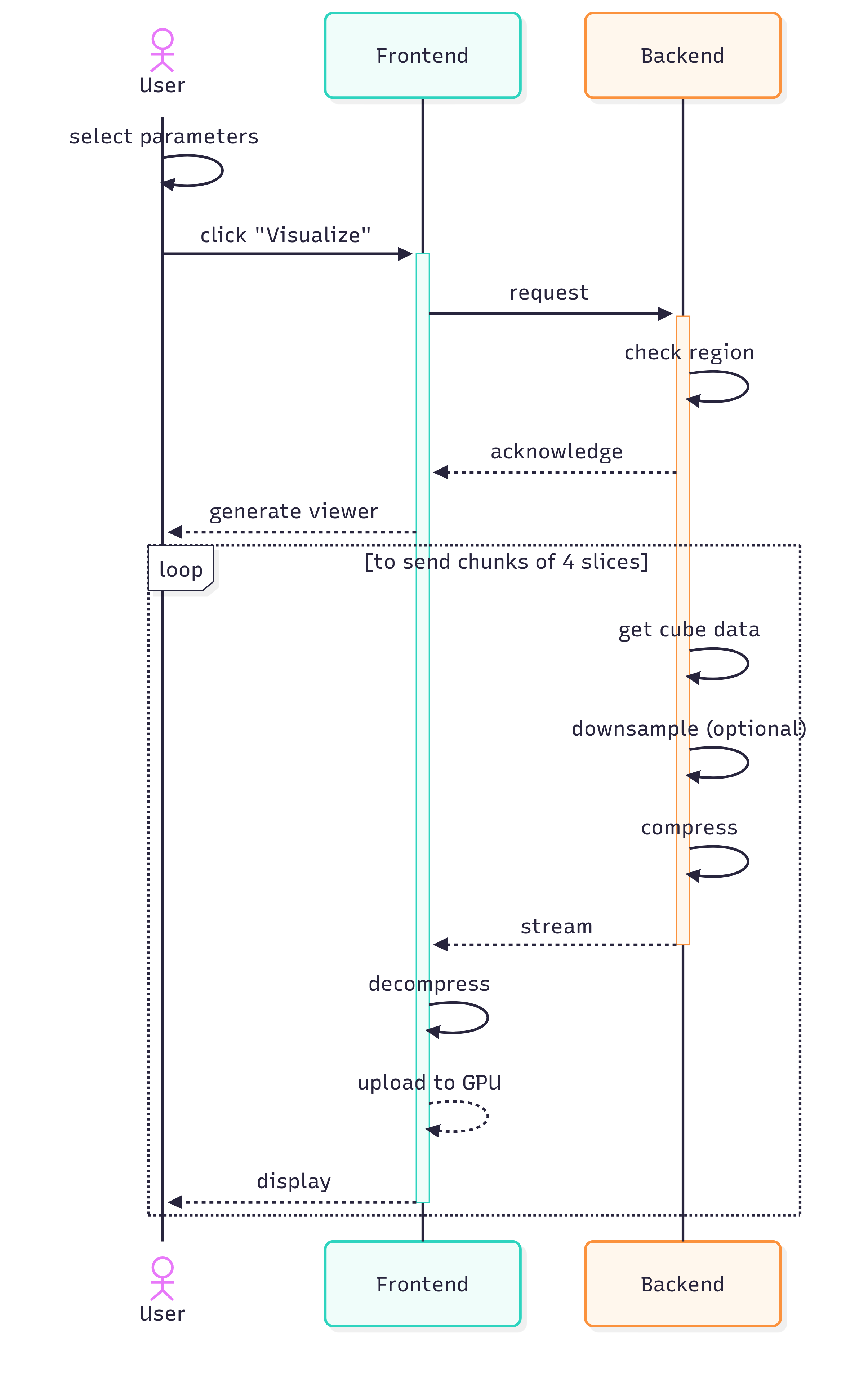}
    \caption{Sequence diagram showing the process to create 3D visualisations including the user interface, frontend and backend.}
    \label{fig:seq_diag}
\end{figure}

The following subsections describe the workflow to produce a 3D visualisation, which also follows the structure described by Figure \ref{fig:architecture}. The workflow includes the request of the visualisation from the frontend to the backend, the streaming of data (sending data in parts) back to the frontend, and rendering. The workflow is shown in the sequence diagram of Figure \ref{fig:seq_diag}.

\subsection{Request}

The 3D rendering widget follows the design of other widgets in CARTA. It is accessible through a toolbar button that opens a panel where users can select the image and region, the spectral range, and downsampling (rebinning) factors for both spatial and spectral dimensions. Clicking the \textit{Visualize} button generates a protobuf message containing the file ID, region ID, viewer ID, spectral range, downsampling factors and compression parameters. This message is sent as a request to the backend. The backend checks that the selected region is valid and responds with an acknowledgement. Upon receiving the acknowledgement, the frontend opens a new panel for the 3D viewer and generates a data store for the required visualisation. The viewer panel initialises an empty ThreeJS 3D texture and an interface to change rendering parameters, such as the minimum and maximum threshold, the colormap or the scale. The data store and the texture are updated as data blocks arrive from the backend.

\subsection{Data stream}

If the backend successfully processes the request, it initiates a stream that sends the data from the requested region in compressed blocks of four slices at a time. Each message in the stream includes the file ID, the region ID, the viewer ID, a compressed subset of the requested data, the corresponding NaN encoding information (location of undefined values in the array, essential for decompression in the frontend), dimensions of the data, index of the block being sent, compression parameters, and a progress indicator.

Initially, CARTA only included 2D compression functions, as existing widgets did not require 3D compression. Therefore, to work with volumetric data, we have developed dedicated functions for 3D compression and NaN encoding. Compression is performed using the ZFP\footnote{\url{https://zfp.io/}} \citep{zfp} algorithm, which is a fixed-rate, error-controlled method optimised for multidimensional arrays. ZFP transforms blocks of $4^{d}$ values with $d$ number of dimensions to a fixed number of bits per block; a high number of bits gives less compression but higher fidelity. This compression scheme was designed for data with spatial correlation, and is more efficient the higher the correlation. Since a 3D $4\times 4\times 4$ block maintains the spatial coherence better than four $4\times 4$ 2D slices, using 3D blocks is more efficient with ZFP. Because the ZFP algorithm does not support NaN values, the NaN encodings function stores indices of NaN values and replaces them in the original data array with the block average. The resulting array, without NaNs, is compressed.

The backend retrieves the 3D datacube using a modified version of CARTA’s \textit{PVpreview} function. We have adapted that code to extract the datacube in blocks of four slices at a time, where the selected 2D axes can be the full image or a smaller region. We have to take downsampling in the spectral axis into account when obtaining the 3D array: the number of slices must be four times the downsampling factor so that we end up with four slices to compress. We have again based our implementation on the \textit{PVpreview} downsampling code. This function first reduces the resolution of individual 2D slices by averaging small regions within the slice. The size of the regions and of the output depends on the spatial downsampling parameter set by the user. Then, a number of slices, dependent on the spectral downsampling parameter, is averaged to obtain blocks of 4 slices. These blocks are the data to be sent to the frontend.

As the data stream reaches the frontend, each block is decompressed using the newly implemented 3D ZFP decompression functions, derived from the existing 2D functions. Decompression consists on reconstructing the 3D array and reintroducing NaN values at their original positions. Because decompression can be computationally demanding, it is executed asynchronously in a web worker.

Decompressed data blocks are sequentially inserted into the corresponding positions of the data store array. The first received block defines the array shape according to the information provided in the message. Each time a new block is processed, a counter in the data store is updated, which in turn triggers an update of the ThreeJS texture that was initialised during the request phase.

\subsection{Rendering}

The texture, even while loading the data blocks, is rendered with a ray marching algorithm based on the open-source code of iDaVIE-v\footnote{\url{https://github.com/idia-astro/iDaVIE/blob/main/}}. In this implementation, we first calculate the volume coordinates to discard rays that do not cross it. Each ray traverses the volume while sampling voxel values in a predefined number of steps. We employ the maximum intensity projection (MIP) method, which consists of finding the maximum value along each path and mapping it to a corresponding colour and opacity. The colour is taken from a chosen colour map after applying set thresholds and scaling functions to the maximum value; the opacity is proportional to that same value. To reduce artifacts that occur when all rays follow the same sampling pattern, known as aliasing artifacts, the starting positions of the rays are initialised with a small randomised offset in the direction perpendicular to the screen: this technique is called ray jittering.

The user can interact with the scene by rotating, zooming or panning it. We have also implemented a simple graphical user interface (GUI) that enables control of the following parameters: the number of steps, adjusting the quality of the rendering; the minimum and maximum thresholds of the data; the colour map; the scale of the colour map; and the scale of the spectral axis. Other parameters could be incorporated in the future.

\begin{figure*}[ht]
    \centering
    \includegraphics[width=\linewidth]{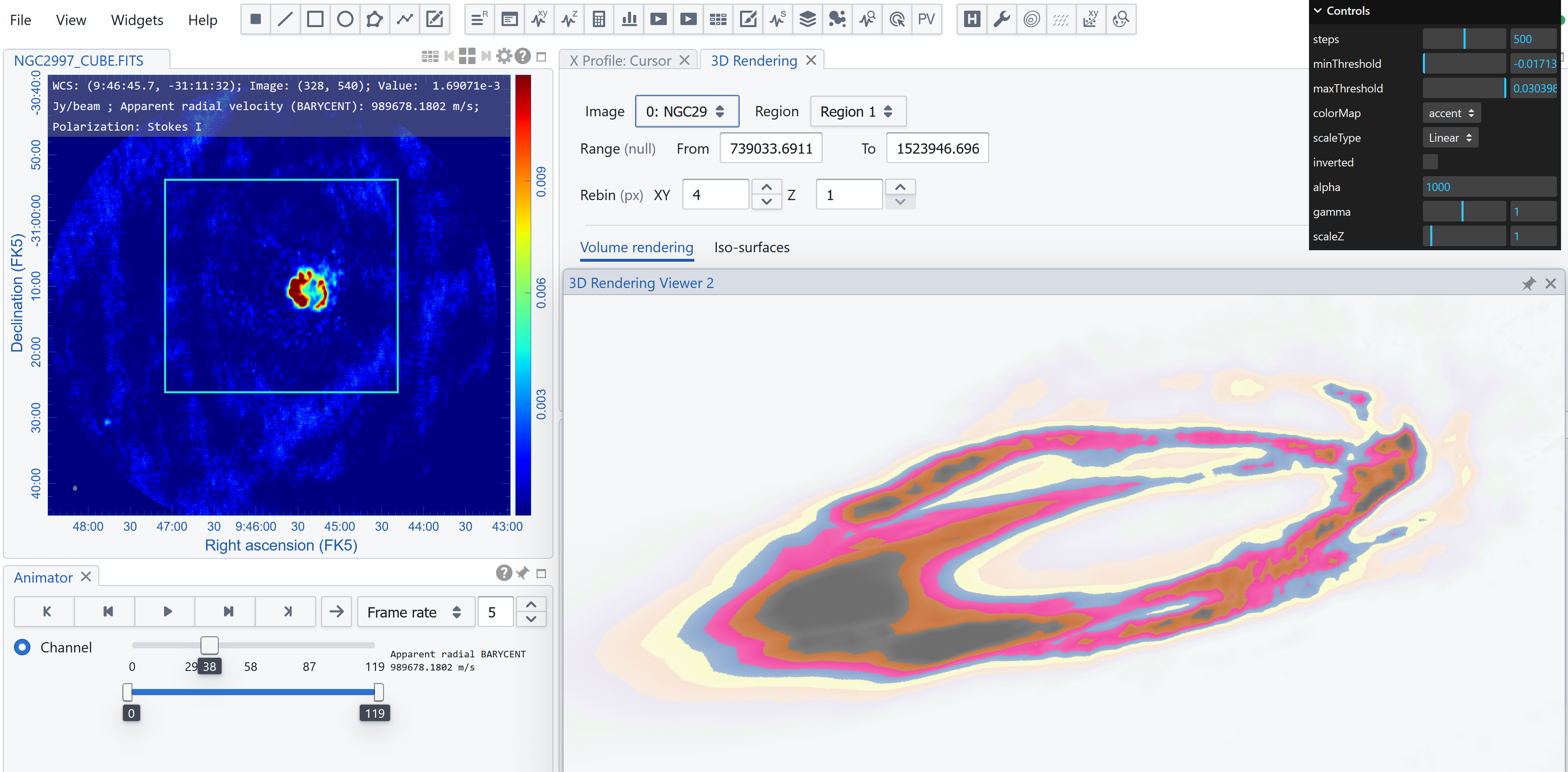}
    \caption{Interface of CARTA showing a 2D view of a datacube slice, the panel of the 3D rendering widget, a 3D visualisation of the selected region, and the interface to interact with the visualisation.}
    \label{fig:interface}
\end{figure*}

CARTA allows multiple concurrent visualisations from different data cubes, limited primarily by available computational resources. This is a powerful capability to compare multiple multi-wavelength observations within a single visual environment. Figure \ref{fig:interface} shows the interface of CARTA within a browser. It includes the 2D view of a single slice of the datacube and the 3D rendering widget with its respective visualisation. We have configured the volume rendering to use the region selected in the 2D view, and chosen the spectral range and the rebinning parameters. In the top right corner we have selected the minimum threshold to be 0, the colour map, and other options.

\section{Results and Discussion}\label{Discussion}

In the previous section we have presented a functional approach to generate 3D visualisations within CARTA. In this section we will discuss the usability of the approach by comparing the effects of different parameters on the performance of the algorithm, both in the frontend and in the backend. We will also mention limitations and possible improvements that could be implemented in the future.

\subsection{Performance}

CARTA is known for its high efficiency and performance with very large datasets: up to 1 TB cubes can be loaded in seconds \citep{hdf52020}. It is important for the 3D rendering widget to match that level of performance and to be integrated seamlessly with the existing data handling structure of CARTA.
We generated a 128 GB datacube ($4000\times4000\times2000$ voxels) containing Gaussian noise and embedded smaller datacubes from real observations. From this cube, we defined subcubes by selecting different spatial regions and different spectral depths. The spatial and spectral ranges were chosen so that certain combinations yield subcubes with the same total number of voxels (e.g., $100 \times 100 \times 400 = 200 \times 200 \times 100 = 4\times10^6$), allowing us to compare cubes with the same volume but different aspect ratios.

We have measured the CPU activity, the RAM usage, and the GPU activity of CARTA during the process of creating a 3D visualisation with different cube sizes and downsampling parameters. The performance tests were conducted on a computer with a 12th Gen Intel(R) Core(TM) i9-12900H processor (14 cores and 20 threads), 32 GB of RAM, and a NVIDIA GeForce RTX 3080 Ti GPU (12 GB of video memory).

\begin{figure}[ht]
    \centering
    \includegraphics[width=\linewidth]{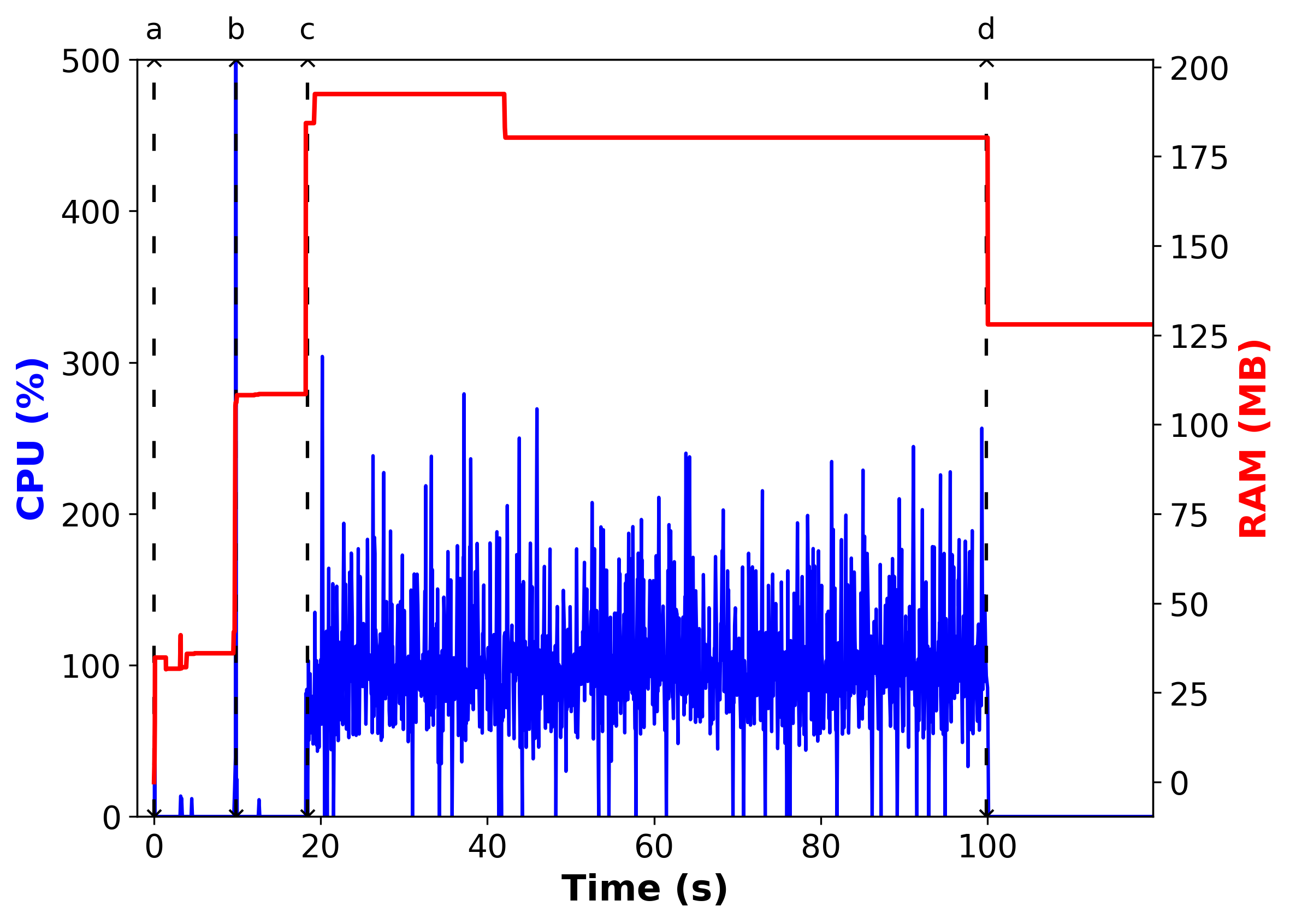}
    \caption{CPU activity (blue) and RAM usage (red) of the backend process during the time required to produce a 3D rendering in CARTA: for a region of $3606\times3605\times1623$ voxels with downsampling factors of 8 and 4 for the spatial plane and spectral axis respectively. The vertical lines represent: a) opening CARTA, b) loading the file, c) requesting the 3D rendering, and d) the end of the data stream. The CPU activity is the sum of 20 threads and could reach 2000\% activity.}
    \label{fig:back_ram_cpu}
\end{figure}

\begin{figure}[ht]
    \centering
    \includegraphics[width=\linewidth]{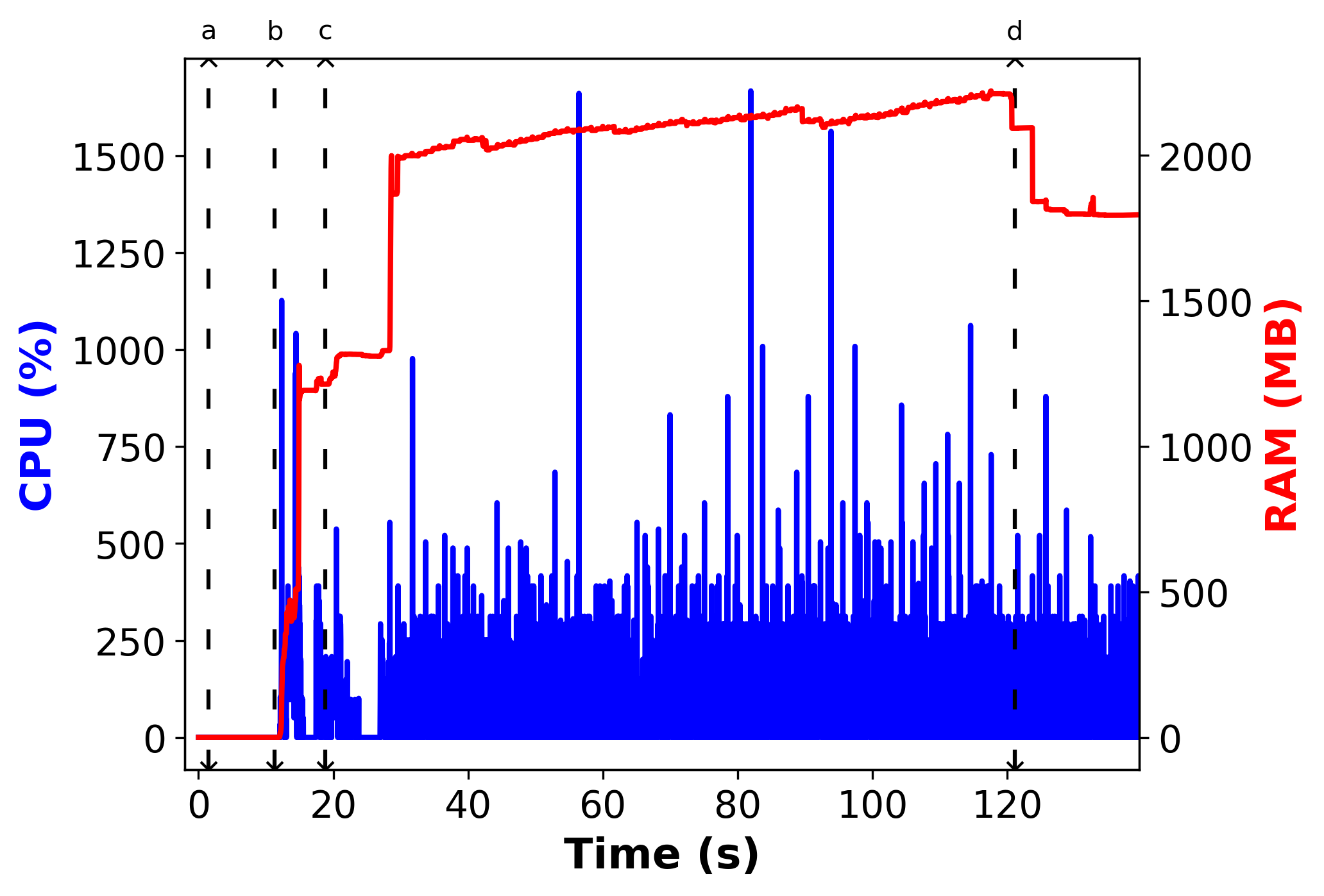}
    \caption{CPU activity (blue) and RAM usage (red) of the frontend process during the time required to produce a 3D rendering in CARTA: for a region of $3606\times3605\times1623$ voxels with downsampling factors of 8 and 4 for the spatial plane and spectral axis respectively. The vertical lines represent: a) opening CARTA, b) loading the file, c) requesting the 3D rendering, and d) the end of the data stream. The CPU activity is the sum of 20 threads and could reach 2000\% activity.}
    \label{fig:front_ram_cpu}
\end{figure}

CPU and RAM profiles of the backend and frontend are presented in Figures \ref{fig:back_ram_cpu} and \ref{fig:front_ram_cpu}, with annotations indicating steps in the process of creating a 3D rendering. The process starts with step a) launching CARTA: the RAM usage increases. Then, in step b) the 128 GB file is loaded: at this instant, the RAM increases again and a large peak is seen in the CPU activity. In step c) the frontend requests a downsampled cube. We see a large increase in the RAM usage and in the CPU activity, which remain high until the streaming of data ends in step d). In the frontend, the overall CPU activity is lower than in the backend, although with some very high peaks (our test machine of 20 threads can reach 2000\%). When all the data is streamed the CPU activity decreases because the rendering is done by the GPU. In most cases the RAM usage in the frontend is higher than in the backend, since, even though the frontend receives downsampled data, all of it must be stored in memory at once. In the backend the data is processed in blocks which usually are smaller than the downsampled datacube. After this point, the CPU activity stays at zero in the backend and the RAM usage decreases.

\begin{figure}[ht]
    \centering
    \includegraphics[width=\linewidth]{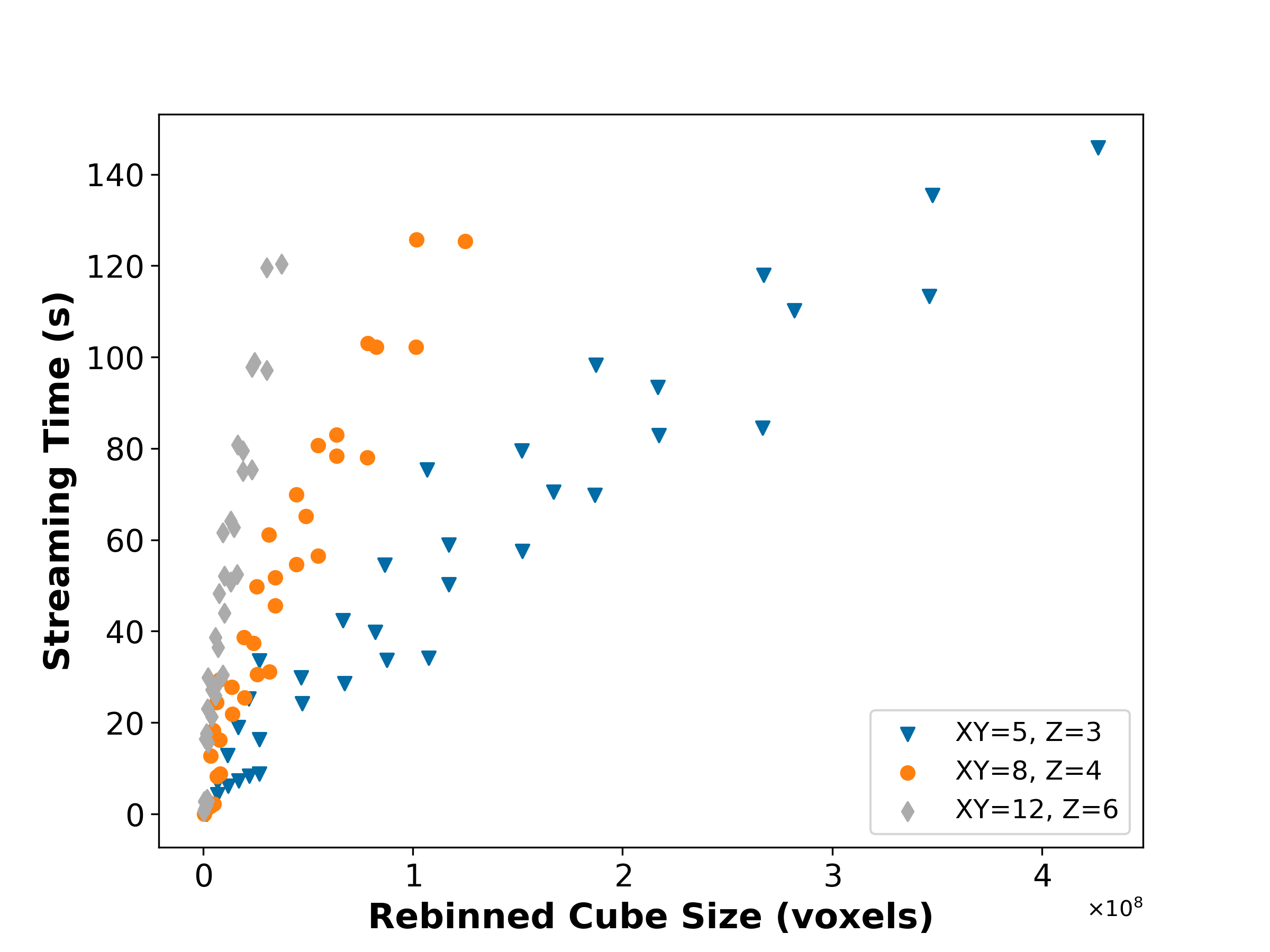}
    \caption{Streaming time against the size of rebinned datacubes: the same region with different rebinning parameters produces different datacube sizes, and different regions with different rebinning parameters can produce the same datacube size. The data is divided into three rebinning parameter combinations.}
    \label{fig:stream_time}
\end{figure}

The time that 3D renderings take to load---the streaming time---has been measured for several regions and is displayed in Figure \ref{fig:stream_time}. The figure shows the time to load subcubes of different sizes for three downsampling parameters. The rebinned cube size is the number of voxels displayed in the visualisation and is obtained by dividing the total number of voxels of the selected original subcube by the rebinning parameters: large rebinned cube size indicates high resolution. The streaming time of the largest cube for each set of parameters ranges from 120 s to 145 s, which is a small increase compared to the change in resolution. This means that using exceedingly high rebinning parameters might not necessary to have a significant effect on loading time.

\begin{figure}[ht]
    \centering
    \includegraphics[width=\linewidth]{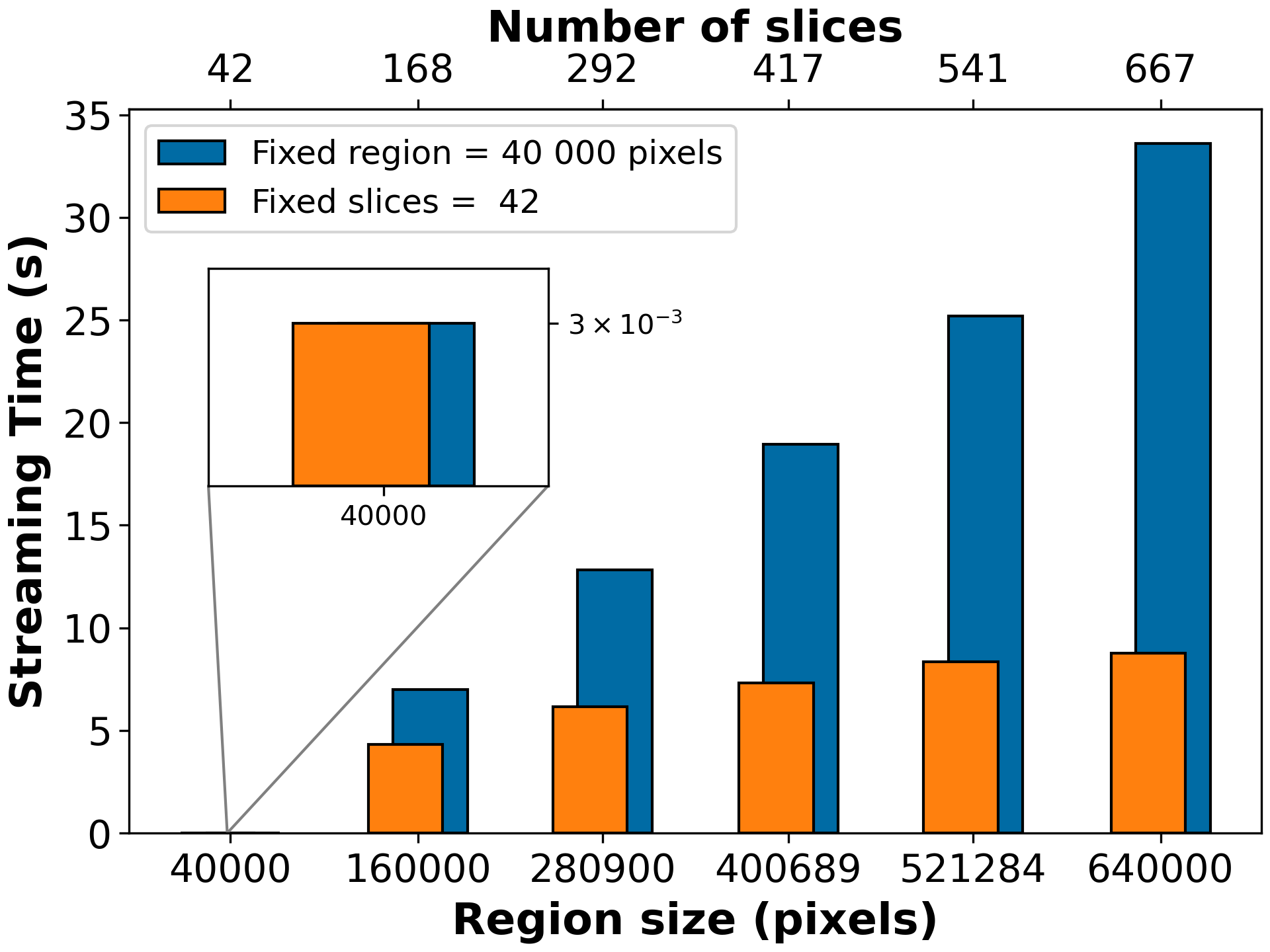}
    \caption{Comparison of the streaming time between pairs of cubes of the same total size but with different numbers of slices and region sizes. The blue bars represent cubes with a fixed region size while the orange bars represent cubes with a fixed number of slices.}
    \label{fig:spatial_spectral}
\end{figure}

\begin{figure}[!ht]
    \centering
    \includegraphics[width=\linewidth]{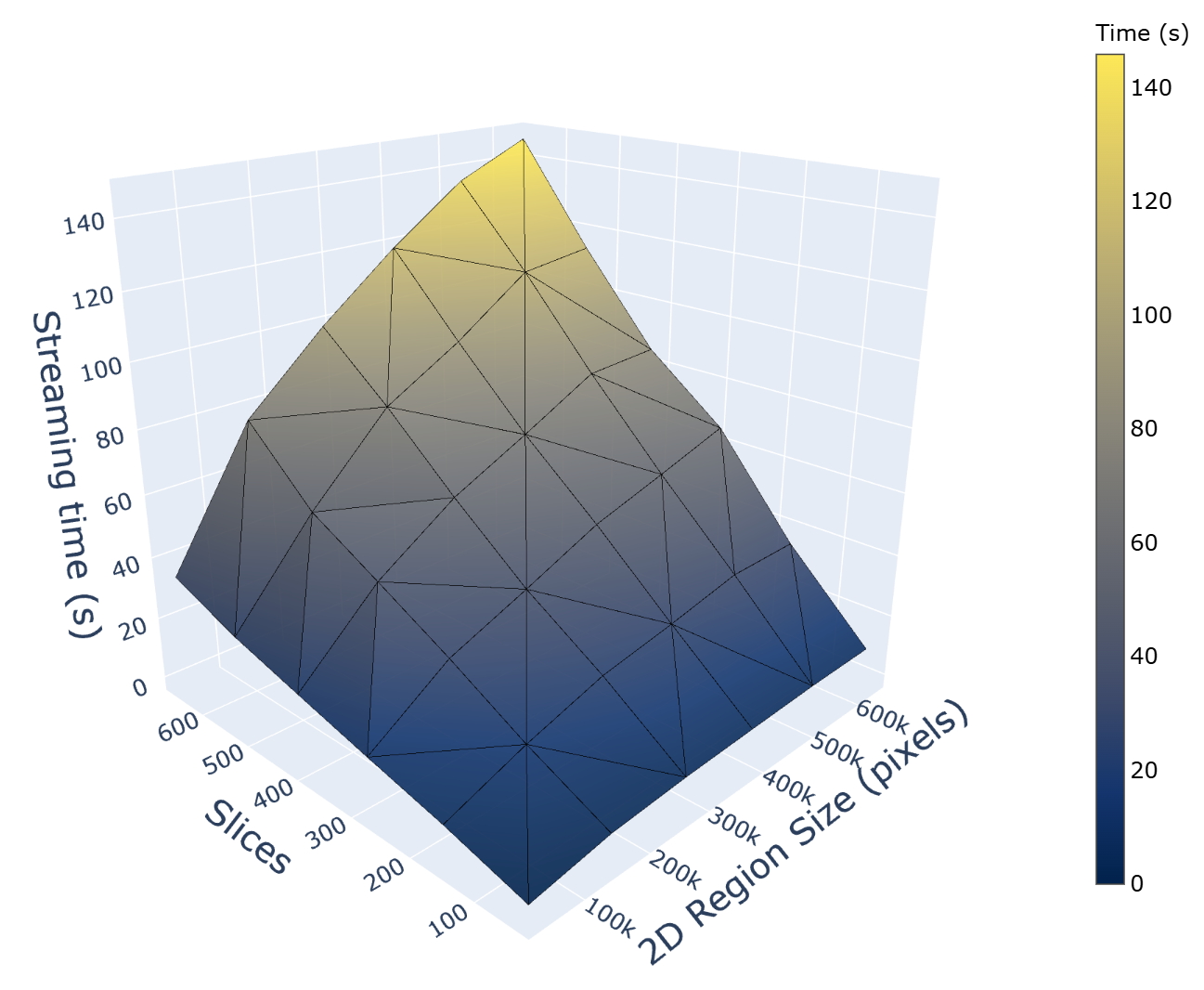}
    \caption{3D plot representing the streaming time needed to load full 3D renderings, as a function of the number of slices in the spectral axis and the size of the 2D region. An interactive visualisation is available as supplementary material.}
    \label{fig:3d_stream}
\end{figure}

\begin{figure*}[ht]
    \centering
    \includegraphics[width=\linewidth]{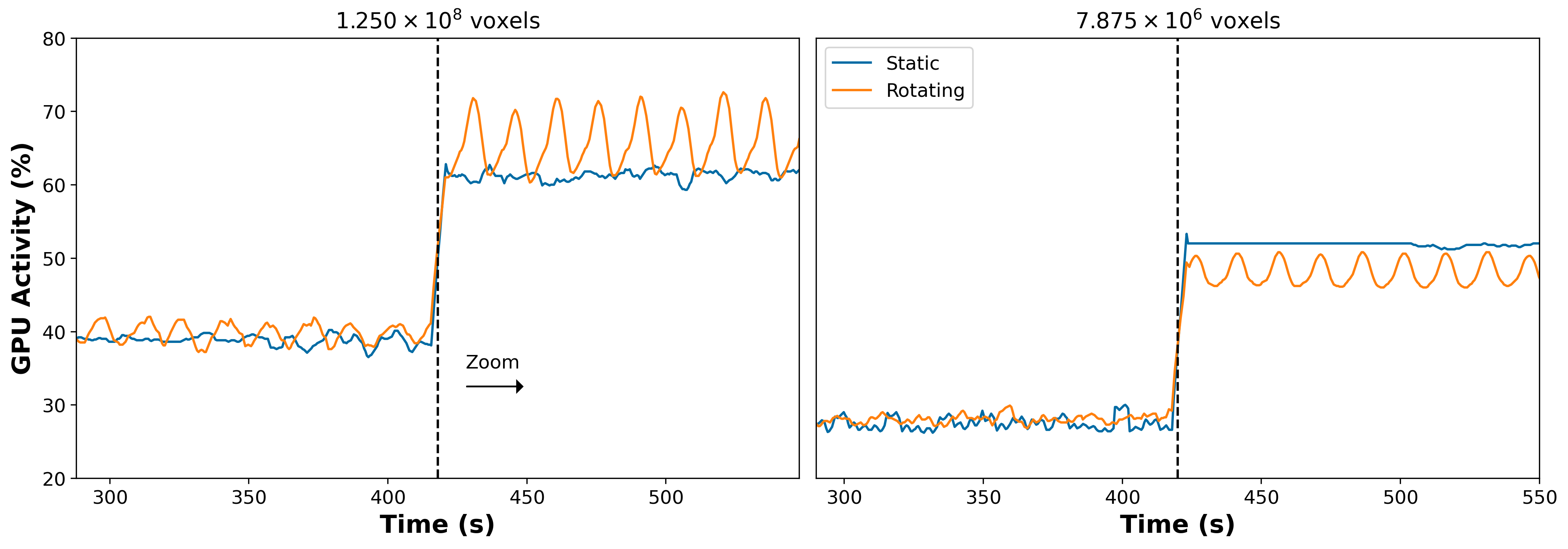}
    \caption{GPU activity through time for two datacubes of different size while they are static or rotating. The left part of each plot was measured with the cube zoomed out, occupying a small part of the screen, and the right part was measured zooming in, with the cube covering the full 3D rendering window. The rotation period is one minute and the sampling rate of the measurements around one second. The display refresh rate was of 240 Hz, which produced 120 fps in the visualisation.}
    \label{fig:gpu_acti}
\end{figure*}

In Figure \ref{fig:spatial_spectral} we show the streaming time for different cube sizes with the same rebinning parameters. The same cube size is obtained in two ways: with a fixed spatial size and with a fixed number of spectral slices. The plot shows that the spectral axis has a larger effect on loading time than the spatial axes. This happens because the datacube is stored as contiguous slices along the spectral axis. As a result, reading a large spatial region across a small number of channels requires only a few sequential reads (data read in order from contiguous storage locations). Conversely, reading a small spatial region across a large number of channels requires repeatedly accessing separated regions of the data, leading to several random reads (data read from non-contiguous storage locations, requiring independent access operations), which are more time consuming than sequential reads. Figure \ref{fig:spatial_spectral} is derived from Figure \ref{fig:3d_stream}, which displays the streaming time for combinations of spatial region sizes and numbers of spectral slices. Similarly to the previous figure, it shows the differing impact of the axes, but it also shows that the streaming time evolves continuously through the parameter space. An interactive 3D plot is delivered as supplementary material.

To assess rendering efficiency, we analyse the GPU activity during the visualisation of fully-loaded datacubes in Figure \ref{fig:gpu_acti}. This figure presents the activity as a function of time for two datacubes of different sizes, showing that rendering a large cube produces a higher activity than rendering a small one. We measured the GPU activity for static renderings and for rotating renderings. We applied rotation in the vertical axis perpendicular to the screen (Y axis) with a period of one minute. We expected a periodic pattern for rotating renderings due to two main factors: variations in ray paths depending on the orientation of the volume and differences in memory access patterns depending on whether sampled data are contiguous. In contrast, static renderings produce nearly constant GPU activity, since rays follow identical paths (except from jittering) and access data in the same direction.

Periodicity is more pronounced when zooming in, as the cube occupies a larger fraction of the screen, as shown in the right parts of Figure \ref{fig:gpu_acti}. The effect of varying path lengths can be mitigated by adapting the step size with the path length, as implemented in, e.g., iDaVIE\footnote{\url{https://github.com/idia-astro/iDaVIE/blob/main/Assets/Shaders/Volumes/BasicVolume.cginc}}. Without zoom (left sides of Figure \ref{fig:gpu_acti}), the activity is still periodic, but this is only observed for the large cube. In this case, both static and rotating cubes present slight variability. The smaller cube shows higher GPU activity while it is static than while it is rotating with the camera positioned inside the volume, which could be explained by the fact that in this position a considerable fraction of the cube is outside the frame: with large cubes this fraction is negligible.

\begin{figure}[ht]
    \centering
    \includegraphics[width=\linewidth]{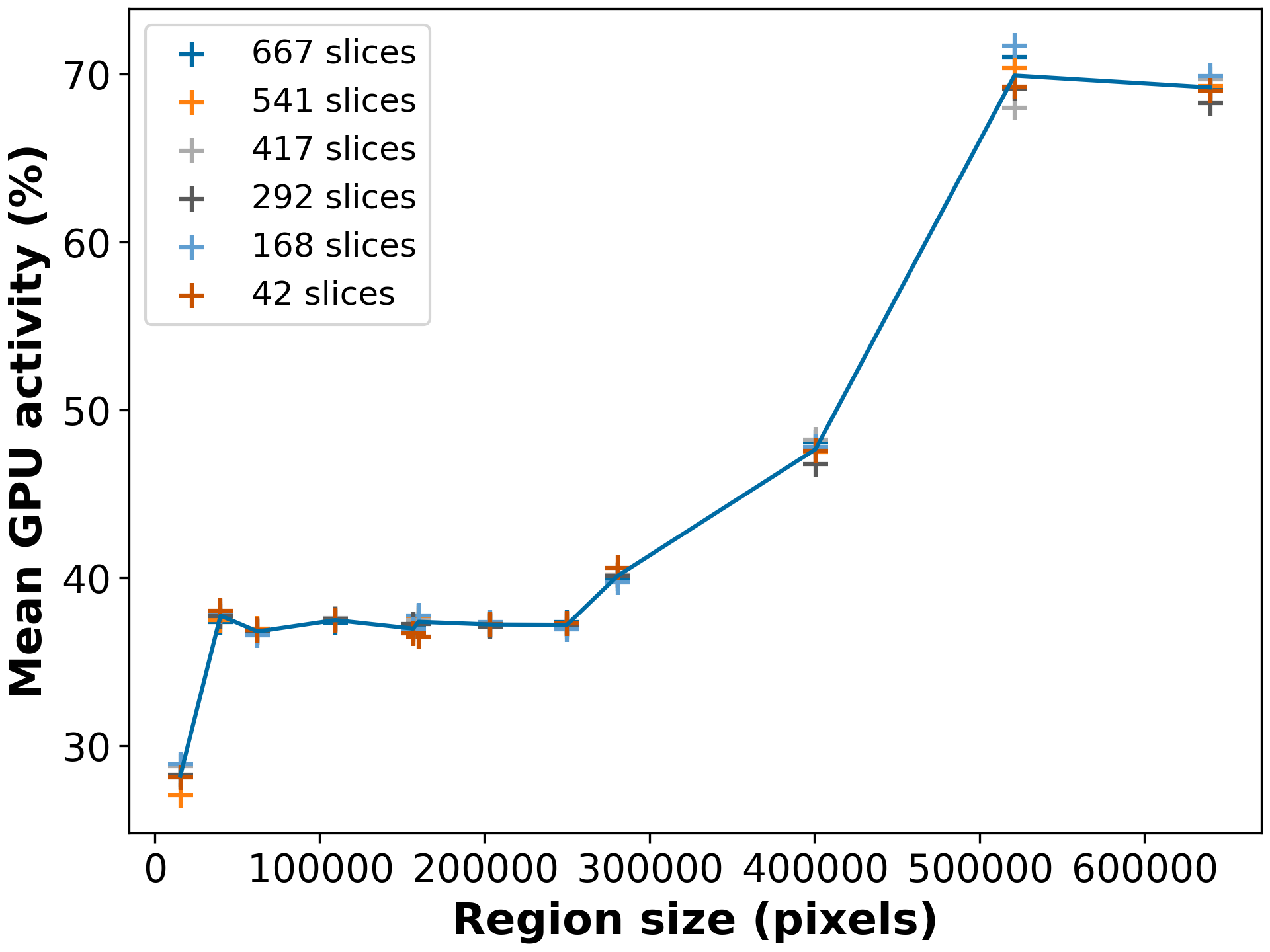}
    \caption{Mean GPU activity against the projected region size of 3D renderings for cubes with different numbers of spectral slices. The line connects the mean values obtained for different numbers of spectral slices corresponding to the same projected 2D region size.}
    \label{fig:gpu_cubesize}
\end{figure}

To further characterise the GPU behaviour, we measured its activity while rendering datacubes of different sizes. Figure \ref{fig:gpu_cubesize} shows the GPU activity as a function of the projected 2D region size for renderings in which the spectral axis is perpendicular to the screen. It indicates that the activity is independent of the depth, as expected because the number of sampling steps per ray is fixed by default. Instead, the activity depends on the projected area, since larger projected regions require more rays to cover the entire cube.

Although the overall trend is an increase in GPU activity with projected area, the relationship is not strictly linear. In particular, the activity remains approximately constant over certain ranges of cube sizes. This behaviour is consistent with GPU occupancy \citep{nvidia_occupancy}, where the processing capacity used by the GPU changes in discrete steps due to processes being allocated in fixed-size blocks. However, other factors may also contribute to the observed behaviour, and more tests, additional profiling tools, and comparisons across different GPU architectures are required for a detailed analysis.

Finally, we have studied the rendering qualitatively. Considering the datacube we have tested (128 GB), once the full datacube is fully loaded, the rendering is displayed at an almost constant frame rate, regardless of the datacube size (we used a display refresh rate of 240 Hz, which produced 120 fps). The rendering can be seen and interacted with while it is still loading blocks of slices. Even so, its usefulness while it is loading depends on the size of the 2D region: large regions need more resources and introducing them in the texture can cause the visualisation to lag. This does not depend on the number of slices. The quality of the rendering, apart from on the downsampling parameters, depends mostly on the number of steps used for ray tracing. We have seen that oversampling---taking more steps than the number of voxels in the longest axis---is needed to obtain high quality rendering.

\subsection{Limitations and improvements}\label{limitations}

We assume that this widget will be used with very large datacubes; therefore it is expected that rendering may take a significant amount of time. As Figure \ref{fig:stream_time} shows, rendering times reach 150 s for our largest test data. We also have to consider that in most cases rendering the whole datacube will not be necessary, since interesting elements will be located in a small region of the image: this will reduce the rendering time.

We found that in our computer, the maximum size of a rendered datacube is limited by the maximum permitted length of a JS array in the browser. That limit can be reached when large datacubes are rendered without downsampling, and this results in an error and an empty volume. The theoretical maximum is calculated from the largest index an array can have, which with 32 bits is $2^{32}-2$ elements. However, because of memory and safety limits, this value is much lower in browsers, depending on the JS engine (e.g., V8, SpiderMonkey) and environment (e.g., Chrome, Firefox). In most cases, the practical value is around 2 GB for the array buffer, which translates to $5\times10^{8}$ elements. There are several solutions or workarounds to this issue: one is setting a limit on the size of the resulting cube and not allowing the request of the visualisation if this limit is passed. Another could be dividing the data into several smaller arrays that do not reach the maximum size. This is a recurrent issue with web-based visualisation tools: \cite{visl3d2025} also find the same problem and fix it by dividing the arrays.

The limitation of volume size in other computers with lower GPU memory could be the maximum WebGL texture size the GPU can allocate or the GPU memory itself. In our case the maximum texture size is $16384\times16384\times2048 \approx 5.5\times10^{11}$ voxels, equivalent to 2 TB in Float32, which is much larger than the available 12 GB of GPU memory. Even so, the GPU memory is larger than the current JS limit and for this reason, the JS error is raised first. This limitation highlights the effectiveness of selecting 2D regions and the spectral range within a cube, which reduces the size of the rendering without loss of information. Downsampling reduces the rendering size at the cost of lower resolution, but can be very useful for a preliminary exploration of a large region.

Regarding possible improvements, adding interactive features is the most feasible in the short term. Creating masks manually is a very convenient capability that is not available in most tools. Masking manually in 3D is useful, for example, to separate diffuse emission from the galaxy. Another option is including catalogues in the 3D rendering, showing markers with positions of galaxies, stars, or other objects. This option already exists in CARTA for the 2D view. iDaVIE is an example of a tool that implements these features successfully. In fact, adding VR capabilities to CARTA could be a major feature that would greatly increase the flexibility of the tool.

In this article, we present how volumetric rendering is included in CARTA; however, there are other visualisation techniques that have other advantages and could be implemented in the future. For example, iso-surface rendering involves calculating surfaces of the same intensity from the data and rendering one or many of them, producing clear boundaries that can be easier to interpret in some cases. Another visualisation technique is point cloud rendering where discrete points are rendered instead of a volume. This method is used mostly for galaxy or star catalogues and in particle simulations, such as cosmological simulations. Introducing vector fields can also be useful, specially if it is combined with other visualisation methods, to display magnetic fields or polarisation.

The user can modify the layout of CARTA, opening and creating various images and configuring different parameters, which can be tedious and take a long time. Most elements of an open CARTA session can be saved to and restored from a workspace, which allows the user to avoid unnecessary repetition of these setup steps. Creating a valid 3D visualisation also requires modifying various parameters (region, thresholds, colormap, scale, etc.) and being able to save it is a convenient feature. In addition, exporting the model into a 3D format such as glTF\footnote{\url{https://www.khronos.org/gltf/}} or X3D\footnote{\url{https://www.web3d.org/}} would increase CARTA's interoperability with other tools, and allow the rendered data to be post-processed in other applications.

\subsection{Design Requirements and Considerations}

Developing a 3D visualisation tool for astronomical data involves balancing multiple design goals, including compliance with FAIR Principles (Findability, Accessibility, Interoperability, and Reusability). These aspects determine not only how easily users can interact with the tool but also how effectively it is integrated into the data analysis ecosystem in astronomy.

Many existing tools for 3D visualisation are technically powerful but often difficult to use. They typically require several pre-processing steps---such as data conversion or parameter configuration---before a 3D rendering can be produced. While such tools offer a wide range of visualisation modes and interactive controls, their complexity results in a steep learning curve that can discourage their use. In this context, usability becomes a critical factor: a visualisation system should be intuitive, responsive, and straightforward, allowing researchers to focus on scientific interpretation rather than on technical setup.

The 3D rendering widget in CARTA addresses this challenge by prioritising simplicity and direct interaction. It can natively open FITS files and generate a 3D visualisation ready to be explored with minimal configuration. Users can easily adjust key visual parameters such as the colormap, intensity thresholds, or scaling, allowing for a balance between automated processing and user control. Moreover, because CARTA already includes analysis and 2D visualisation tools, integrating 3D rendering within the same environment allows scientists to transition seamlessly between qualitative and quantitative data exploration.

Increasing the interoperability of CARTA could extend this ability to transition to datasets or analysis tools. For example, the Virtual Observatory (VO) provides a global framework of standards and protocols designed to enable seamless interaction among datasets, services, and other resources from different facilities. Integrating VO functionalities into CARTA would enable it to access multi-wavelength data sources, query distributed archives, and contribute to interoperable workflows. In particular, the 3D rendering widget would benefit from this by adding functionalities to download catalogues or images from the VO and overlaying them in the 3D visualisation, similar to how VisIVO or ViSL3D do.

Scalability and performance are factors that visualisation applications must take into account because of the large scale of data to be handled---especially from new observatories such as SKAO. CARTA's client–server architecture addresses this issue, allowing heavy data processing to be performed server-side while maintaining an interactive experience for the user. The use of other efficient techniques, such as the HDF5 schema, enhances its performance. In addition, CARTA can be deployed in a Docker container, which makes it highly portable and reproducible. Combining CARTA containers with management software such as Kubernetes adds on scalability, for example, by providing separate containers on-demand for multi-user access.

\section{Conclusions}\label{conclusions}

The exploration of astronomical datacubes requires tools able to create 3D visualisations that effectively display all the information present in the data. The large scale of data available, which will increase in the future, presents a challenge for current 3D visualisation tools, requiring the exploration of different approaches to offer visualisation services. CARTA is a powerful tool prepared for astronomy that includes several analysis features as well as 2D visualisation options. Since a lot of data products from radio observatories and from IFUs are datacubes, enabling the creation of 3D visualisations will greatly increase the utility of the application.

In this article, we have presented an approach to implement a 3D rendering widget in CARTA, taking advantage of its client-server architecture to make it scalable for big data. The widget requests a part or the whole datacube from the backend and renders it as a volume in the frontend. This way, users maintain control of various aspects of the visualisation, such as the colour, the scale, and others; while being able to render large datasets locally. In addition, the compression of data, sending it in blocks, and downsampling algorithms increase the efficiency of the process.

We have studied the performance of the widget by testing it with a 128 GB datacube. We have requested visualisations for different combinations of 2D regions, spectral ranges, and downsampling parameters and measured the CPU activity, the RAM usage, and the GPU activity of both the frontend and the backend when running CARTA locally. The tests show that our approach succeeds in creating effective 3D renderings from large datacubes, and highlights the advantage of downsampling or selecting regions within the cube to be able to visualise them.

We have seen that loading time depends more on the size of the spectral axis than of the spatial axes, and that while using smaller downsampling parameters increases loading time, the difference is not substantial in practice and very large downsampling is not needed in most cases. The RAM required to produce a visualisation increases while the blocks of data are streamed: this increase is the sum of the size of individual spectral slices, which indicates that RAM usage depends on the size of the spatial region, not the size of the spectral axis.

Our approach paves the way for an addition of a 3D visualisation widget into CARTA; even so, more development is needed to fully include it in a release. Other improvements, such as including catalogues, could be added in the future, but they are not essential for inclusion of the widget in the software. CARTA is already efficient, user friendly, and has many applications, but adding other features, such as VO access, scripting capabilities, or the 3D rendering widget presented in this article, would greatly increase its value for the astronomical community.

\section*{Acknowledgements}

Authors ILG, MPR, SSE, LVM, and JG acknowledge financial support from the grant PID2021-123930OB-C21 and PID2024-155817OB-I00 funded by MICIU/AEI/ 10.13039/501100011033 and by ERDF/EU, the grant CEX2021-001131-S funded by MICIU/AEI/ 10.13039/501100011033, and the grant TED2021-130231B-I00 funded by MICIU/AEI/ 10.13039/501100011033 and by the European Union NextGenerationEU/PRTR. ILG also acknowledges financial support from PRE2021-100660 funded by MICIU/AEI/ 10.13039/501100011033 and by ESF+.

%% Authors AC, AP, and MvZ acknowledge financial support from...


%\section{Appendix title 1}

\bibliographystyle{elsarticle-harv}
\bibliography{bibliography}

@ARTICLE{hdf52020,
       author = {{Comrie}, A. and {Pi{\'n}ska}, A. and {Simmonds}, R. and {Taylor}, A.~R.},
        title = "{Development and application of an HDF5 schema for SKA-scale image cube visualization}",
      journal = {Astronomy and Computing},
         year = 2020,
        month = jul,
       volume = {32},
          eid = {100389},
        pages = {100389},
          doi = {10.1016/j.ascom.2020.100389},
       adsurl = {https://ui.adsabs.harvard.edu/abs/2020A&C....3200389C}
}

@misc{srcnet0.1_devplan,
    author = {{Salgado}, Jesús and {Joshi}, Rohini and {Sánchez-Expósito},
    Susana and {Parra-Royón}, Manuel and {Walder}, James and {Llopis}, Pablo and
    {Guo}, Shaoguang and  {Morris}, Dave and  {Taffoni}, Giuliano and 
    {Gheller}, Claudio and {Ouellette}, John and {Kang}, Hyunwoo and 
    {Gaudet}, Séverin and {An}, Tao and {Akahori}, Takuya and {Swinbank},
    John and {Hess}, Kelley Michelle and {Oonk}, Raymond and {Kok}, Tim},
    title = "{SRCNet v0.1 Implementation Plan}",
    year = 2024,
    month = jun,
    url = "https://confluence.skatelescope.org/display/SNC/SRCNet+Documents"
}

@article{hassanfluke2011_3Dreview, 
    title={Scientific Visualization in Astronomy: Towards the Petascale Astronomy Era}, 
    volume={28}, DOI={10.1071/AS10031}, 
    number={2}, 
    journal={Publications of the Astronomical Society of Australia}, 
    publisher={Cambridge University Press}, 
    author={Hassan, Amr and Fluke, Christopher J.}, 
    year={2011}, 
    pages={150–170}
}

@ARTICLE{Lan2021_3Dreview,
       author = {{Lan}, Fangfei and {Young}, Michael and {Anderson}, Lauren and {Ynnerman}, Anders and {Bock}, Alexander and {Borkin}, Michelle A. and {Forbes}, Angus G. and {Kollmeier}, Juna A. and {Wang}, Bei},
        title = "{Visualization in Astrophysics: Developing New Methods, Discovering Our Universe, and Educating the Earth}",
      journal = {arXiv e-prints},
         year = 2021,
        month = may,
          eid = {arXiv:2106.00152},
        pages = {arXiv:2106.00152},
          doi = {10.48550/arXiv.2106.00152},
archivePrefix = {arXiv},
       eprint = {2106.00152},
 primaryClass = {astro-ph.IM},
       adsurl = {https://ui.adsabs.harvard.edu/abs/2021arXiv210600152L}
}

@ARTICLE{visivo2015,
       author = {{Sciacca}, E. and {Becciani}, U. and {Costa}, A. and {Vitello}, F. and {Massimino}, P. and {Bandieramonte}, M. and {Krokos}, M. and {Riggi}, S. and {Pistagna}, C. and {Taffoni}, G.},
        title = "{An integrated visualization environment for the virtual observatory: Current status and future directions}",
      journal = {Astronomy and Computing},
         year = 2015,
        month = jun,
       volume = {11},
        pages = {146-154},
          doi = {10.1016/j.ascom.2015.01.006},
       adsurl = {https://ui.adsabs.harvard.edu/abs/2015A&C....11..146S}
}

@ARTICLE{visl3d2025,
       author = {{Labadie-Garc{\'\i}a}, I. and {Garrido}, J. and {Verdes-Montenegro}, L. and {Mendoza}, M. {\'A}. and {Parra-Roy{\'o}n}, M. and {S{\'a}nchez-Exp{\'o}sito}, S. and {Ianjamasimanana}, R.},
        title = "{3D radio data visualisation in open science platforms for next-generation observatories}",
      journal = {Astronomy and Computing},
         year = 2025,
        month = jul,
       volume = {52},
          eid = {100949},
        pages = {100949},
          doi = {10.1016/j.ascom.2025.100949},
archivePrefix = {arXiv},
       eprint = {2503.16237},
 primaryClass = {astro-ph.IM},
       adsurl = {https://ui.adsabs.harvard.edu/abs/2025A&C....5200949L}
}

@INPROCEEDINGS{idavie2024,
       author = {{Jarrett}, Thomas and {Comrie}, Angus and {Sivitilli}, Alexander and {Pretorius}, Pieter Cilliers and {Vitello}, Fabio and {Marchetti}, Lucia},
        title = "{iDaVIE: Immersive Data Visualisation Interactive Explorer}",
    booktitle = {Zenodo Software},
         year = 2024,
       volume = {46},
        month = oct,
    publisher = {Zenodo},
          eid = {4614115},
        pages = {4614115},
          doi = {10.5281/zenodo.4614115},
       adsurl = {https://ui.adsabs.harvard.edu/abs/2024zndo...4614115T}
}

@misc{carta,
  author       = {Angus Comrie and
                  Kuo-Song Wang and
                  Yu-Hsuan Hwang and
                  Adrianna Pińska and
                  Pamela Harris and
                  Carli Raul-Omar and
                  Hou, Kuan-Chou and
                  Aikema, David and
                  Cheng-Chin Chiang and
                  Ming-Yi, Lin and
                  Huang, Po-Sheng and
                  Gao, Zhen-Kai and
                  Rob Simmonds},
  title        = {CARTA: The Cube Analysis and Rendering Tool for
                   Astronomy
                  },
  month        = aug,
  year         = 2025,
  publisher    = {Zenodo},
  version      = {5.0},
  doi          = {10.5281/zenodo.17050846},
  url          = {https://doi.org/10.5281/zenodo.17050846},
}

@INPROCEEDINGS{meerkat2016,
       author = {{Jonas}, J. and {MeerKAT Team}},
        title = "{The MeerKAT Radio Telescope}",
    booktitle = {MeerKAT Science: On the Pathway to the SKA},
         year = 2016,
        month = jan,
          eid = {1},
        pages = {1},
          doi = {10.22323/1.277.0001},
       adsurl = {https://ui.adsabs.harvard.edu/abs/2016mks..confE...1J}
}

@ARTICLE{ALMA2009,
       author = {{Wootten}, Alwyn and {Thompson}, A. Richard},
        title = "{The Atacama Large Millimeter/Submillimeter Array}",
      journal = {IEEE Proceedings},
         year = 2009,
        month = aug,
       volume = {97},
       number = {8},
        pages = {1463-1471},
          doi = {10.1109/JPROC.2009.2020572},
archivePrefix = {arXiv},
       eprint = {0904.3739},
 primaryClass = {astro-ph.IM},
       adsurl = {https://ui.adsabs.harvard.edu/abs/2009IEEEP..97.1463W}
}

@INPROCEEDINGS{muse2010,
       author = {{Bacon}, R. and {Accardo}, M. and {Adjali}, L. and {Anwand}, H. and {Bauer}, S. and {Biswas}, I. and {Blaizot}, J. and {Boudon}, D. and {Brau-Nogue}, S. and {Brinchmann}, J. and {Caillier}, P. and {Capoani}, L. and {Carollo}, C.~M. and {Contini}, T. and {Couderc}, P. and {Daguis{\'e}}, E. and {Deiries}, S. and {Delabre}, B. and {Dreizler}, S. and {Dubois}, J. and {Dupieux}, M. and {Dupuy}, C. and {Emsellem}, E. and {Fechner}, T. and {Fleischmann}, A. and {Fran{\c{c}}ois}, M. and {Gallou}, G. and {Gharsa}, T. and {Glindemann}, A. and {Gojak}, D. and {Guiderdoni}, B. and {Hansali}, G. and {Hahn}, T. and {Jarno}, A. and {Kelz}, A. and {Koehler}, C. and {Kosmalski}, J. and {Laurent}, F. and {Le Floch}, M. and {Lilly}, S.~J. and {Lizon}, J. -L. and {Loupias}, M. and {Manescau}, A. and {Monstein}, C. and {Nicklas}, H. and {Olaya}, J. -C. and {Pares}, L. and {Pasquini}, L. and {P{\'e}contal-Rousset}, A. and {Pell{\'o}}, R. and {Petit}, C. and {Popow}, E. and {Reiss}, R. and {Remillieux}, A. and {Renault}, E. and {Roth}, M. and {Rupprecht}, G. and {Serre}, D. and {Schaye}, J. and {Soucail}, G. and {Steinmetz}, M. and {Streicher}, O. and {Stuik}, R. and {Valentin}, H. and {Vernet}, J. and {Weilbacher}, P. and {Wisotzki}, L. and {Yerle}, N.},
        title = "{The MUSE second-generation VLT instrument}",
    booktitle = {Ground-based and Airborne Instrumentation for Astronomy III},
         year = 2010,
       editor = {{McLean}, Ian S. and {Ramsay}, Suzanne K. and {Takami}, Hideki},
       series = {Society of Photo-Optical Instrumentation Engineers (SPIE) Conference Series},
       volume = {7735},
        month = jul,
          eid = {773508},
        pages = {773508},
          doi = {10.1117/12.856027},
archivePrefix = {arXiv},
       eprint = {2211.16795},
 primaryClass = {astro-ph.IM},
       adsurl = {https://ui.adsabs.harvard.edu/abs/2010SPIE.7735E..08B},
}

@ARTICLE{jwst2006,
       author = {{Gardner}, Jonathan P. and {Mather}, John C. and {Clampin}, Mark and {Doyon}, Rene and {Greenhouse}, Matthew A. and {Hammel}, Heidi B. and {Hutchings}, John B. and {Jakobsen}, Peter and {Lilly}, Simon J. and {Long}, Knox S. and {Lunine}, Jonathan I. and {McCaughrean}, Mark J. and {Mountain}, Matt and {Nella}, John and {Rieke}, George H. and {Rieke}, Marcia J. and {Rix}, Hans-Walter and {Smith}, Eric P. and {Sonneborn}, George and {Stiavelli}, Massimo and {Stockman}, H.~S. and {Windhorst}, Rogier A. and {Wright}, Gillian S.},
        title = "{The James Webb Space Telescope}",
      journal = {\ssr},
         year = 2006,
        month = apr,
       volume = {123},
       number = {4},
        pages = {485-606},
          doi = {10.1007/s11214-006-8315-7},
archivePrefix = {arXiv},
       eprint = {astro-ph/0606175},
 primaryClass = {astro-ph},
       adsurl = {https://ui.adsabs.harvard.edu/abs/2006SSRv..123..485G}
}

@ARTICLE{askap2021,
       author = {{Hotan}, A.~W. and {Bunton}, J.~D. and {Chippendale}, A.~P. and {Whiting}, M. and {Tuthill}, J. and {Moss}, V.~A. and {McConnell}, D. and {Amy}, S.~W. and {Huynh}, M.~T. and {Allison}, J.~R. and {Anderson}, C.~S. and {Bannister}, K.~W. and {Bastholm}, E. and {Beresford}, R. and {Bock}, D.~C. -J. and {Bolton}, R. and {Chapman}, J.~M. and {Chow}, K. and {Collier}, J.~D. and {Cooray}, F.~R. and {Cornwell}, T.~J. and {Diamond}, P.~J. and {Edwards}, P.~G. and {Feain}, I.~J. and {Franzen}, T.~M.~O. and {George}, D. and {Gupta}, N. and {Hampson}, G.~A. and {Harvey-Smith}, L. and {Hayman}, D.~B. and {Heywood}, I. and {Jacka}, C. and {Jackson}, C.~A. and {Jackson}, S. and {Jeganathan}, K. and {Johnston}, S. and {Kesteven}, M. and {Kleiner}, D. and {Koribalski}, B.~S. and {Lee-Waddell}, K. and {Lenc}, E. and {Lensson}, E.~S. and {Mackay}, S. and {Mahony}, E.~K. and {McClure-Griffiths}, N.~M. and {McConigley}, R. and {Mirtschin}, P. and {Ng}, A.~K. and {Norris}, R.~P. and {Pearce}, S.~E. and {Phillips}, C. and {Pilawa}, M.~A. and {Raja}, W. and {Reynolds}, J.~E. and {Roberts}, P. and {Roxby}, D.~N. and {Sadler}, E.~M. and {Shields}, M. and {Schinckel}, A.~E.~T. and {Serra}, P. and {Shaw}, R.~D. and {Sweetnam}, T. and {Troup}, E.~R. and {Tzioumis}, A. and {Voronkov}, M.~A. and {Westmeier}, T.},
        title = "{Australian square kilometre array pathfinder: I. system description}",
      journal = {\pasa},
         year = 2021,
        month = mar,
       volume = {38},
          eid = {e009},
        pages = {e009},
          doi = {10.1017/pasa.2021.1},
archivePrefix = {arXiv},
       eprint = {2102.01870},
 primaryClass = {astro-ph.IM},
       adsurl = {https://ui.adsabs.harvard.edu/abs/2021PASA...38....9H}
}

@INPROCEEDINGS{aladin2022,
       author = {{Baumann}, Matthieu and {Boch}, Thomas and {Pineau}, Fran{\c{c}}ois-Xavier and {Fernique}, Pierre and {Bot}, Caroline and {Allen}, Mark},
        title = "{Aladin Lite v3: Behind the Scenes of a Major Overhaul}",
    booktitle = {Astronomical Data Analysis Software and Systems XXX},
         year = 2022,
       editor = {{Ruiz}, Jose Enrique and {Pierfedereci}, Francesco and {Teuben}, Peter},
       series = {Astronomical Society of the Pacific Conference Series},
       volume = {532},
        month = jul,
        pages = {7},
       adsurl = {https://ui.adsabs.harvard.edu/abs/2022ASPC..532....7B}
}

@misc{nvidia_occupancy,
  author       = {{NVIDIA Corporation}},
  title        = {CUDA C++ Best Practices Guide: Occupancy},
  year         = {2025},
  edition      = {Version 13.3},
  url = {https://docs.nvidia.com/cuda/cuda-c-best-practices-guide/index.html#occupancy},
  note         = {Accessed: 2026-07-10}
}

@article{marching2003,
    author = {Thomas {Lewiner} and Hélio {Lopes} and Antônio {Wilson Vieira} and Geovan {Tavares}},
    title = {Efficient Implementation of Marching Cubes' Cases with Topological Guarantees},
    journal = {Journal of Graphics Tools},
    volume = {8},
    number = {2},
    pages = {1-15},
    year = {2003},
    publisher = {Taylor & Francis},
    doi = {10.1080/10867651.2003.10487582},
    URL = {https://doi.org/10.1080/10867651.2003.10487582},
    eprint = {https://doi.org/10.1080/10867651.2003.10487582}
}

@ARTICLE{zfp,
  author={Lindstrom, Peter},
  journal={IEEE Transactions on Visualization and Computer Graphics}, 
  title={Fixed-Rate Compressed Floating-Point Arrays}, 
  year={2014},
  volume={20},
  number={12},
  pages={2674-2683},
  doi={10.1109/TVCG.2014.2346458},
}

@INPROCEEDINGS{ds92003,
       author = {{Joye}, W.~A. and {Mandel}, E.},
        title = "{New Features of SAOImage DS9}",
    booktitle = {Astronomical Data Analysis Software and Systems XII},
         year = 2003,
       editor = {{Payne}, H.~E. and {Jedrzejewski}, R.~I. and {Hook}, R.~N.},
       series = {Astronomical Society of the Pacific Conference Series},
       volume = {295},
        month = jan,
        pages = {489},
       adsurl = {https://ui.adsabs.harvard.edu/abs/2003ASPC..295..489J}
}

\end{document}